\documentstyle[epsfig]{article}

\def\tou#1{{\lower1.2ex\hbox{$\longrightarrow$}\atop
        {\lower-.7ex\hbox{$\scriptscriptstyle #1 $}}}}
\def\lsim{{\lower1.2ex\hbox{$<$}\atop
        {\lower-.7ex\hbox{$\sim$}}}}
\def\gsim{{\lower1.2ex\hbox{$>$}\atop
        {\lower-.7ex\hbox{$\sim$}}}}

\def\be{\begin{equation}}
\def\ee{\end{equation}}

\begin{document}

\begin{titlepage}
\rightline {Si-95-04 \  \  \  \   }

\vspace*{2.truecm}

\centerline{\Large \bf  Finite Size Scaling and Critical Exponents}
\vskip 0.6truecm
\centerline{\Large \bf
in Critical Relaxation }
\vskip 0.6truecm

\vskip 2.0truecm
\centerline{\bf Zhibing Li$^{\star}$\footnote {On leave of absence
from Zhongshan University,
510275 Guangzhou, P.R. China}, Lothar Sch\"ulke, and Bo Zheng}
\vskip 0.2truecm
\centerline{$^{\star}$ I.C.T.P., I -- 34100 Trieste, Italy}

\vskip 0.2truecm
\centerline{Universit\"at -- GH Siegen, D -- 57068 Siegen, Germany}

\vskip 2.truecm

\abstract{
We simulate the critical relaxation process of the
two-dimensional Ising model with the initial state
both completely disordered or completely ordered.
   Results of a new method to measure both the dynamic and static
	critical exponents are reported, based on
	the finite size scaling for the dynamics at the early time.
{}From the time-dependent Binder cumulant, the
dynamical exponent $z$ is extracted independently, while the
static exponents $\beta/\nu$ and $\nu$ are obtained
from the time evolution of the magnetization and its
higher moments.
}


\end{titlepage}

\section{Introduction}

   For statistical systems in equilibrium or
near equilibrium critical phenomena arise around the second order
phase transition points. Due to the infinite spatial and time
correlation lengths there appear universality and scaling.
The universal behaviour of critical systems is characterized by
the critical exponents. The determination of critical exponents
has long been one of the main interests for both analytical
calculations and
 numerical simulations.

   Numerically critical exponents are usually measured by
 generating the configurations in the
equilibrium with Monte Carlo methods. To obtain the critical
exponents from the finite size scaling,
   Binder's method is widely accepted \cite{bin92,bin81}.
   The dynamical exponent $z$ is
traditionally measured from the exponential decay of the
time correlation for finite systems in the long-time regime
 \cite{wan91,wil85}.
   As is well known, numerical simulations
near the critical point suffer from critical slowing
down. Much effort
has been made to circumvent this difficulty.
To study the static properties of the system,
some {\it non-local} algorithms, e.g., the cluster algorithm
\cite{swe87,wol89}, have proved to be very efficient compared to
the normal {\it local} algorithms. However, in this case
the original dynamic
universality class is altered by the non-locality of the algorithm.
Properties of the original local dynamics can not be obtained with
non-local algorithms.

In recent years the exploration of critical phenomena
has been broadened. Universality and scaling are also
discovered for systems
far from equilibrium. Better understanding has been achieved
of the critical relaxation process even up to the early time.
   A representative example for such a process is
that the Ising model initially in a random state with
a small magnetization
is suddenly quenched to the critical temperature and
then evolves according to the dynamics of model A.
Janssen, Schaub and Schmittmann \cite{jan89} have
argued by an $\epsilon$-expansion up to two--loop order that,
besides the well known universal  behaviour in the long-time
regime, there exists
another {\it universal} stage of the relaxation
{\it at early times}, the
so-called {\it critical initial slip}, which sets in right
after the
microscopic time scale $t_{mic}$.
The characteristic time scale for the critical initial slip
is $t_0 \sim m_0^{-z/x_0}$, where $m_0$ is the initial
magnetization and
$x_0$ is the dimension of it. It has been shown that $x_0$ is
a new
 independent critical exponent for describing the critical
dynamic system.

The characteristic behaviour of the critical initial slip
is that, when a non-zero initial magnetization $m_0$ is generated,
due to the anomalous dimension of the operator $m_0$
the time dependent magnetization $M(t)$ undergoes a critical
initial increase
\begin{equation}
M(t) \sim m_0 \, t^\theta,
\label{cis}
\end{equation}
where $\theta$ is related to $x_0$ by $x_0 =  \theta z + \beta/\nu$.
The exponent
$\theta$ has been measured with Monte Carlo simulation
for the Ising model
and the Potts model
both directly from the power law increase of the magnetization
in (\ref {cis}) \cite {li94,sch95} and
indirectly from the power law decay of the auto-correlation
\cite{hus89}-\nocite{hum91}\cite{men94}. The results are in good
agreement with those
from  an $\epsilon$-expansion and the scaling relation is
confirmed.

In a previous paper \cite{li95}
we proposed to measure both the dynamic and static
exponents from the finite size scaling of the dynamic relaxation
at the early time. The idea is demonstrated for the 2-dimensional
Ising model. Since the measurement is carried out
from the beginning of the time evolution, the method is efficient
at least for the dynamic exponent $z$. Even though certain
aspects of this dynamic approach  should still
be clarified, the results indicate a possible broad application
of the short-time dynamics
since the universal behaviour of the dynamics at early time
is found to be quite general
\cite{oer93}-\nocite{oer94}\nocite{bra94}\nocite{rit95}
\nocite{bra95}\cite{gra95}.

One of the purposes of the present paper is
to give a detailed and complete
analysis of the data briefly reported in \cite{li95}.
While in that letter we have extracted the exponents by the
optimal fit of two curves in a certain time interval
(``global'' fit), we propose here in addition a new approach by which
the critical exponents are obtained for each time step separately
(``local'' fit).
Furthermore the
simulation has been extended to a longer time interval in order
to confirm the stability of the measurements
in the time direction and to see how the scaling possibly
passes over to the long-time regime.

On the other hand, we may easily realize that
for the scaling
of the short-time dynamics a small initial magnetization is
important   besides the short
initial correlation length.
This is because $m_0=0$ is a fixed point for the
renormalization group transformation.
However, there exits also another fixed point
corresponding to $m_0=1$.
Therefore one may like to know whether around that fixed point
universality and scaling are
also  present or not.
Actually some trials have been made with Monte Carlo
simulation \cite{sta92,mun93}.
For a large enough lattice, one may expect a power
law decay of the magnetization
\be
M(t) \sim t^{-\frac{\beta}{\nu z}}
\ee
before the exponential decay starts.  From this behaviour
the exponent $\beta/{(\nu z)}$ can be estimated.
However the results are not yet complete.

Therefore another purpose of this paper is
to present a systematic investigation
of the finite size scaling for the critical relaxation
starting from $m_0=1$. It will be shown
that scaling is observed in the early stage of the time
evolution and with the help of the finite size scaling
all the critical exponents $z$, $\beta$ and $\nu$
can be obtained from the lattices which are much smaller
that those in \cite{sta92,mun93}.

The following section~2 is devoted to
the critical relaxation with $m_0=0$ and Sect.~3 to $m_0=1$.
The final section contains some discussion.

\section{The Critical Relaxation \protect\newline with Zero
         Initial Magnetization}

The Hamiltonian for the Ising model is
\begin{equation}
H=J  \sum_{<ij>}  S_i\ S_j\;,\qquad S_i=\pm 1\;,
\label{hami}
\end{equation}
with $<ij>$ representing nearest neighbours.
In the equilibrium the Ising model is exactly solvable.
The critical point locates at $J_c=\log(1+\surd 2)/2$,
and the exponents $\beta=1/8$ and $\nu=1$ are known.
In principle any type of the dynamics can be given to the system
to study the non-equilibrium evolution processes.
Unfortunately up to now none of them can be solved
exactly.

In this paper we consider only the dynamics of model A.
For the numerical simulation, typical examples are
the Monte Carlo Heat-Bath algorithm and the Metropolis
algorithm. For the analytical calculation, the Ising model
should be assumed to be described
by the $\lambda \phi ^4$ theory. Then the Langevin equation
can be introduced as a dynamic equation.
For the Langevin dynamic system the renormalization group method
may be applied to understand the critical behaviour
as universality and scaling.
For the critical relaxation with the initial condition
of a very short correlation and small magnetization,
Janssen, Schaub and Schmittmann \cite{jan89} have performed a
perturbative renormalization calculation
with an $\epsilon$-expansion
up to two--loop order. They have obtained the scaling relation
which is valid even
in the short-time regime, and all
the critical exponents including the new dynamic exponent
$\theta$ which governs the initial behaviour of the critical
relaxation.

Of special interest is here
the extension of the scaling form in Ref.~\cite{jan89} to
finite-size systems \cite{die93,li94}.
In accordance to the renormalization group analysis
for finite-size
systems,
after a microscopic time scale $t_{mic}$ we expect
a scaling relation to hold for the k-th moment of the
magnetization in the neighbourhood of the critical point
starting from  the macroscopic short-time regime
\cite{pri84,jan89,zin89},
\begin{equation}
M^{(k)}(t,\tau,L,m_{0})=b^{k\beta/\nu}
M^{(k)}(b^{z}t,b^{-1/\nu}\tau,bL,
b^{-x_{0}}m_{0})
\label{e2}
\end{equation}
assuming that the initial correlation length is zero
 and the initial magnetization $m_0$ is small enough.
Here $t$ is the dynamic evolution time,
$\tau=(T-T_{c})/T_{c}$ is the reduced temperature,
$L$ is the lattice size,
and $b$ is the spatial rescaling factor.
It has been discussed that under certain conditions
the effect of $m_0$ remains even in the long-time regime
of the critical relaxation \cite {rit95a}. This modifies the
traditional scaling relation where the effect of $m_0$
has usually been suppressed.

In this paper we are only interested in the measurement
of the well-known critical exponents $z$, $\beta$, and $\nu$.
To make the computation simpler and more efficient,
we set $m_{0}$ to its
fixed point $m_{0}=0$ in this section.
Therefore the exponent $x_{0}$
will not enter
the calculation. Furthermore, now the time scale
$t_{0}=m_0^{-z/x_0} \to \infty$, and the
critical initial slip gets most prominent in time direction
even though the magnetization itself will only fluctuate
around zero.

The initial state  with $m_0=0$
is prepared by starting from a lattice with all spins equal, then
the spins of randomly chosen sites are switched,
until exactly half of the spins are opposite.
This initial state is updated with the Heat-Bath algorithm to
 $300$ time steps for $L=8, 16, 32$, and to
 $900$ time steps for $L=64$.
The average over 50\,000 samples of this kind
with independent initial configurations has been taken in each run,
and 8 runs are used to estimate the errors.
The critical value $J_c=0.4406$ has been used and, in order
to fix $1/\nu$ separately, we have repeated all simulations with
$J=0.4386$.
In each case the observables $|M(t)|$,
$ M^{(2)}(t)$ and $M^{(4)}(t)$
have been measured.

To determine $z$ {\it independently}, we introduce a
{\it time-dependent}
 Binder cumulant
\be
U(t,L) =1-\frac{ M^{(4)} }{ 3 (M^{(2)})^2}
\label{ez}
\ee
Here the argument $\tau$ has been set to zero and skipped.
 \
The simple finite size scaling relation
\be
U(t,L)=U(t',L');\qquad t' = b^z t, \qquad L' = b L
\label{e4}
\ee
for the cumulant
is easily deduced from Eq.(\ref{e2}).

The exponent $z$ can easily be obtained through
searching for a time scaling factor $b^{z}$ such that
 the cumulants from two different lattices
in both sides in Eq.(\ref{e4}) collapse.
We call this global scaling fitting.

%############################## BEGINN  TAB1 ####################
\begin{table}[h]\centering
$$
\begin{array}{|c|c|c|c|l|l|l|}
\hline
\mbox{Input} & \mbox{Lattice} & t'_{min} & t'_{max}  &\quad z\ &
\quad 2\beta/\nu\ &\quad 1/\nu\  \\
\hline
\hline
           &\ 8\leftrightarrow 16 & \ 10 & 300  & 2.100(2) &
           0.2473(02) & 1.13(1)  \\
\cline{5-7}
           & 16\leftrightarrow 32 & \ 10 &  &  2.149(2) &
           0.2494(06) & 1.08(4)  \\
\cline{2-7}
 U          &                   & \ 50 & 300  & 2.134(4) &
 0.2510(11) & 1.00(5)  \\
\cline{5-7}
         &   32\leftrightarrow 64  &  200 &   & 2.140(4) &
         0.2488(11) & 1.03(4)  \\
\cline{3-7}
 M^2          &                  & \ 50 & 900 & 2.151(2) &
 0.2531(08) & 1.02(2)  \\
\cline{5-7}
          &                      &  200 &  &  2.153(2) &
          0.2523(08) & 1.02(2)  \\
\hline
\hline
 \tilde U  & 32\leftrightarrow 64 & \ 50 & 900 & 2.151(3) &
 0.2515(11) & 1.03(2)  \\
\cline{5-7}
|M|        &                      &  200 &  &2.152(3) &
0.2521(11) & 1.03(2)  \\
\hline
\end{array}
$$
\caption{
\footnotesize
 Results for $z$, $2\beta/\nu$ and $1/\nu$, respectively,
from the 2-dimensional Ising model with initial magnetization $m_0=0$.
The values are
obtained from a global scaling fit for two lattices.
}
\label{t1}
\end{table}

%############################## ENDE TAB1 #########################
%############################## BEGINN  TAB2  #####################
\begin{table}[t]\centering
$$
\begin{array}{|c|c|c|c|l|l|l|}
\hline
\mbox{Input} & \mbox{Lattice} &  t'_{min} & t'_{max} &\quad z\ &
\quad 2\beta/\nu\ &\quad
1/\nu\  \\
\hline
\hline
           &\ 8\leftrightarrow 16 & \ 10 & 300 &  2.333(4) &
           0.2372(08) &
1.11(1)  \\
\cline{2-7}
 U        & 16\leftrightarrow 32 & \ 10 & 300 &  2.143(2) &
 0.2448(11) &
1.19(5)  \\
\cline{2-7}
 M^2          & 32\leftrightarrow 64 & \ 50 & 900 &  2.148(3) &
 0.2510(16) &
1.07(2)  \\
\cline{5-7}
          &                      &  200 &    &  2.155(2) &
          0.2488(10) &
1.05(2)  \\
\hline
\hline
 \tilde U  & 32\leftrightarrow 64 & \ 50 & 900 &  2.144(4) &
 0.2541(28) &
1.07(2)  \\
\cline{5-7}
|M|        &                      &  200 &     &  2.153(3) &
0.2501(14) &
1.05(2)  \\
\hline
\end{array}
$$
\caption{
\footnotesize
 Results for $z$, $2\beta/\nu$ and $1/\nu$, respectively,
from the $2$-dimensional Ising model with initial magnetization
$m_0=0$.
Values are from an
average in time direction with a local scaling fit.
}
\label{t2}
\end{table}
%############################## ENDE TAB2  ######################

\begin{figure}[p]\centering
\epsfysize=12cm
\epsfclipoff
\fboxsep=0pt
\setlength{\unitlength}{1cm}
\begin{picture}(13.6,12)(0,0)
\put(0.2,8.2){\makebox(0,0){$U(t,L)$}}
\put(10.7,.3){\makebox(0,0){$t$}}
\put(0,0){{\epsffile{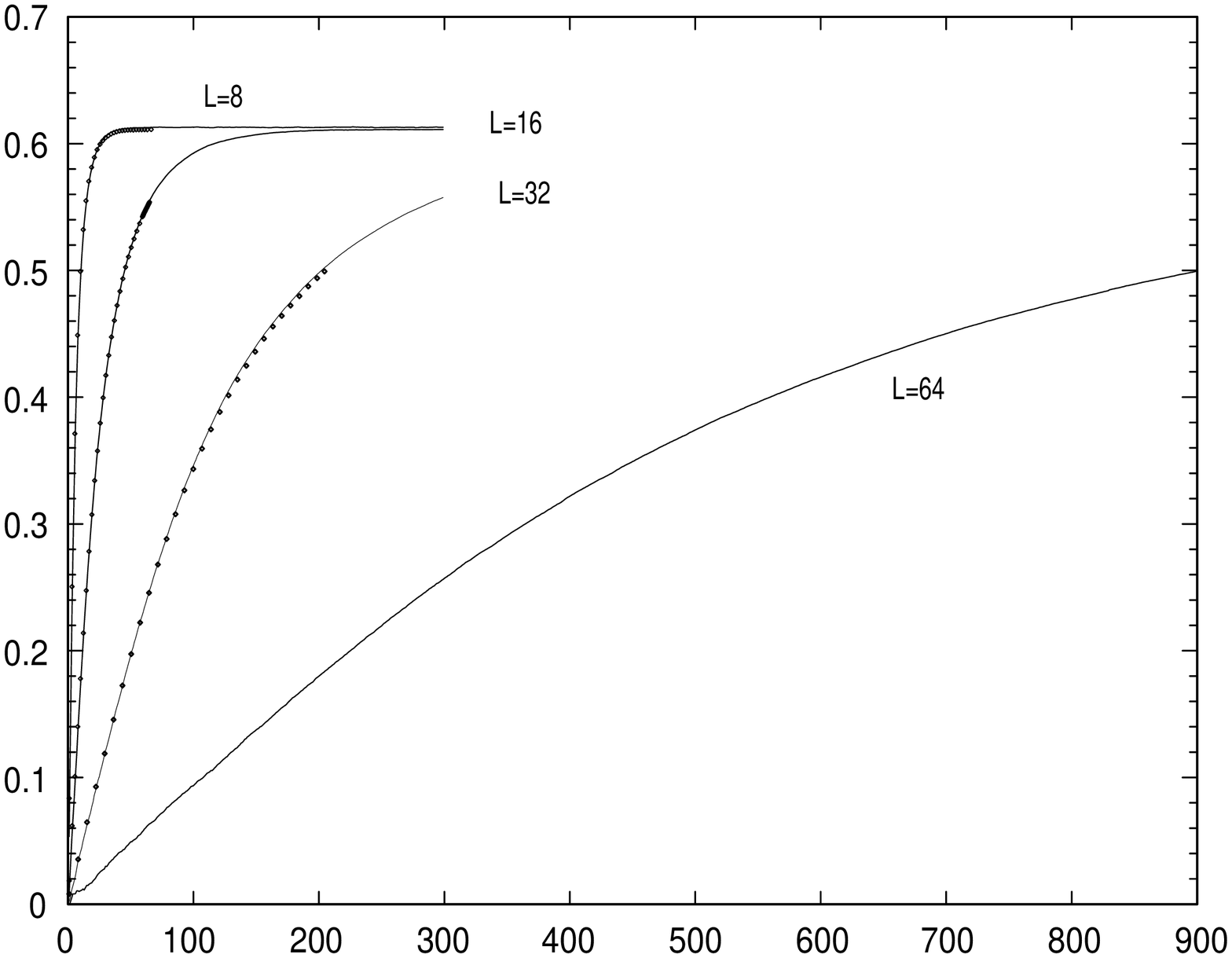}}}
\end{picture}
\caption{\footnotesize
	The cumulants $U(t,L)$ for $L=8$, $16$, $32$ and $64$
for initial magnetization $m_0=0$
	plotted
	versus the time $t$. The dots fitted to the lines show
	the cumulants with lattice size $2L$ rescaled in time
        by the best fit value $2^z$ given in Table~1.
}
\label{ddU}
\end{figure}

Here the cumulant $U(t,L)$ obtained from each of the 8 runs
for lattice size $L$
has been compared with each run for $L'=2L$.  The best scaling
factor $2^{z}$ has
been estimated by the method of least squares.
Figure~\ref{ddU} shows the cumulants for $L=8$,$16$,$32$ and $64$ by
solid lines. The dots fitted to the lines for $L$ up to $32$ show the
results for $L'=2L$ rescaled in time
 with the best fitting scaling factor $2^z$.
 Only a selected number of 30 equidistant points has been plotted.
One sees the remarkable scaling collapse even in the short-time
regime. Compared with the figure shown
in the previous paper \cite{li95}
the evolution time for $L'=64$ has been extended up to
$t'_{max}=900$.
The average value for $z$ and
the error estimated from this procedure have been given in
Tab.~\ref{t1}
for different pairs of lattices using Eq. (\ref{e4}).
In the first steps of the time evolution, the
values of $M^{(2)}$ and $M^{(4)}$
are quite small.
A careful view at the data shows that
their accuracy and in particular the accuracy of the cumulant
$U(t,L)$ is not so good as for lager $t$. One may expect
that skipping the region of smaller t will give more reliable
results.
Therefore we have performed fits for
different time intervals $[t'_{min},t'_{max}]$.
The results in reference \cite{li95} corresponds to
$t'_{min}=1$ and $t'_{max}=300$. Fig.~\ref{ddU} is from
a fit with $t'_{min}=50$ and $t'_{max}=900$.
The longer evolution time of $t'_{max}=900$ for lattice
$L=64$ shows also the stability of the
scaling in the time direction. From the results
we can see that $z$ for larger time $t$ is slightly bigger
than that for smaller time $t$.
Later we will come back to this point.

\begin{figure}[p]\centering
\epsfysize=12cm
\epsfclipoff
\fboxsep=0pt
\setlength{\unitlength}{1cm}
\begin{picture}(13.6,12)(0,0)
\put(0.0,8.2){\makebox(0,0){$M^{(2)}(t,L)$}}
\put(10.8,.3){\makebox(0,0){$t$}}
\put(0,0){{\epsffile{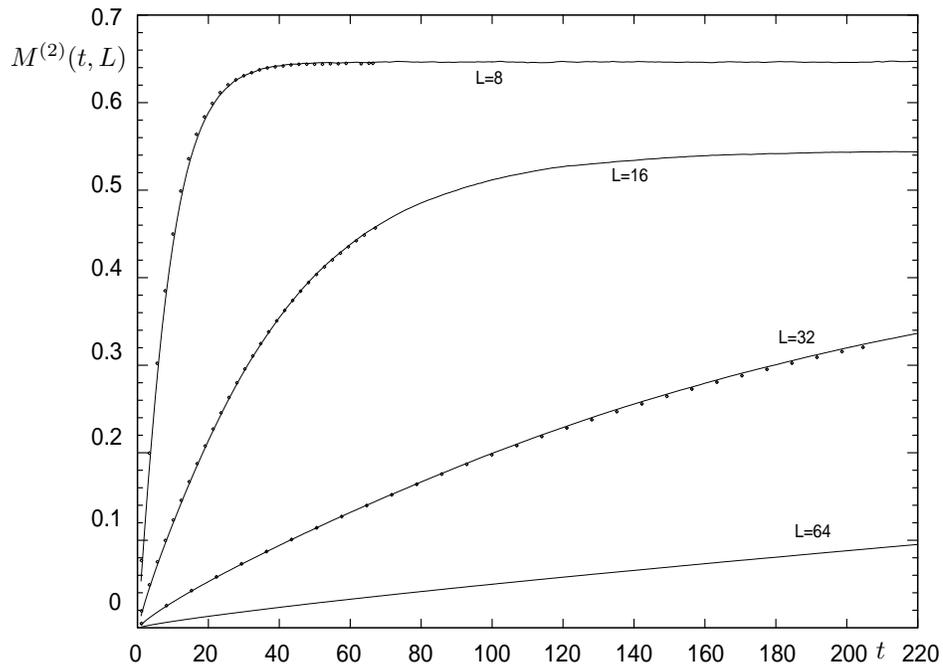}}}
\end{picture}
\caption{\footnotesize
	The second moments $M^{(2)}(t,L)$ for $L=8$, $16$, $32$ and $64$
for initial magnetization $m_0=0$
	plotted
	versus the time $t$. The dots fitted to the lines show
	the second moments with lattice size $2L$ rescaled in time
         by the best fit value $2^z$ and
	$2\beta/\nu$ given in Table~1.
}
\label{ddM}
\end{figure}

With $z$ in hand, the scaling relation for the second moment
\be
M^{(2)}(t,L)=b^{2\beta/\nu} M^{(2)}(t',L');
\qquad t'=b^{z}t \qquad L'=bL,
\label{e5}
\ee
can be used to estimate the exponent $2\beta/\nu$ in a similar way.
The results have been included in Tab.~\ref{t1}.
The curves for
$M^{(2)}$ for the different lattice sizes and the corresponding
scaling fits
can be found in Fig.~\ref{ddM}. We have cut the time scale
at $t=220$ in order to show the data relevant for the
scaling fit more clearly.

\begin{figure}[p]\centering
\epsfysize=12cm
\epsfclipoff
\fboxsep=0pt
\setlength{\unitlength}{1cm}
\begin{picture}(13.6,12)(0,0)
\put(-.3,8.1){\makebox(0,0){$\partial_\tau\ln M^{(2)}(t,L)$}}
\put(10.8,.3){\makebox(0,0){$t$}}
\put(0,0){{\epsffile{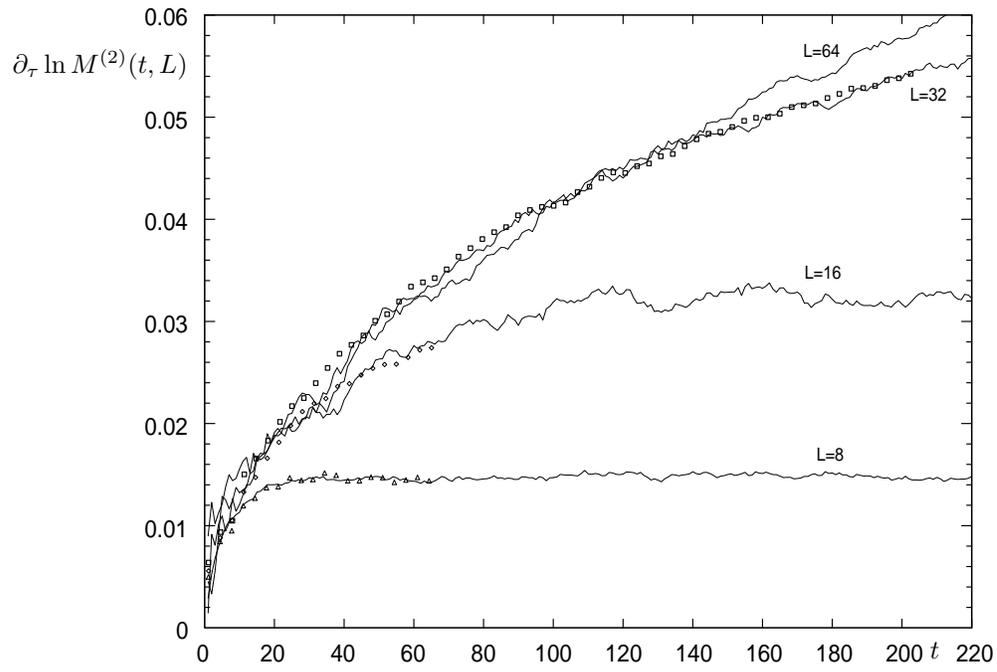}}}
\end{picture}
\caption{\footnotesize
        $\partial_\tau\ln M^{(2)}(t,L)$ for $L=8$, $16$, $32$
	and $64$
for initial magnetization $m_0=0$
	plotted
	versus the time $t$. The dots fitted to the lines show
	those with lattice size $2L$ rescaled in time
	by the best fit values $2^z$ and
	$1/\nu$ given in Table~1.
}
\label{ddd}
\end{figure}

Slightly more complicated appears the determination of
$1/\nu$. One can use the derivative with respect to $\tau$ either
of $U$ or of $M^{(2)}$. Here the latter gives
more stable results. From
\be
\partial_\tau \ln M^{(2)}(t,\tau,L)|_{\tau=0}
=
b^{-{1/\nu}} \partial_{\tau'} \ln M^{(2)}(t',\tau',L')|_{\tau'=0}
\label{es}
\ee
with $t'=b^{z}t$ and $L'=b L$, the exponent $1/\nu$ can
independently be
calculated. The derivative is approximated by
taking the difference of the
values for $M^{(2)}$ at $J=J_c=0.4406$ and $J=0.4386$, divided
by its value at $J_c$.
It is clear and can also be seen in Fig.~\ref{ddd} that the
result for this related quantity fluctuates more than that
for $U(t,L)$, in particular for small $t$.
The results of this calculation have also been included
in the last column of Tab.~\ref{t1}.

It is interesting to point out, see Tab.~\ref{t1}, that
the results from the scaling fit of $L=16$ and $L=32$ are already
quite good.
This is probably due to the fact that the spatial
correlation length in the short-time regime of the dynamic
evolution is very small
and therefore from small lattices one can already obtain
reasonable results.
We also like to
mention that the procedure of comparing each run for the small
lattice with each run for the big lattice may
underestimate the errors. Therefore the data including the
errors given in
Tab.~\ref{t1} sometimes do not cover the exact values.
The real errors may be a factor two bigger.

\begin{figure}[p]\centering
\epsfysize=12cm
\epsfclipoff
\fboxsep=0pt
\setlength{\unitlength}{1cm}
\begin{picture}(13.6,12)(0,0)
\put(12.5,8.6){\makebox(0,0){$z=2.14$}}
\put(12.5,4.4){\makebox(0,0){$1/\nu=1.0$}}
\put(12.7,1.5){\makebox(0,0){$2\beta/\nu=0.25$}}
\put(10.8,.3){\makebox(0,0){$t$}}
\put(0,0){{\epsffile{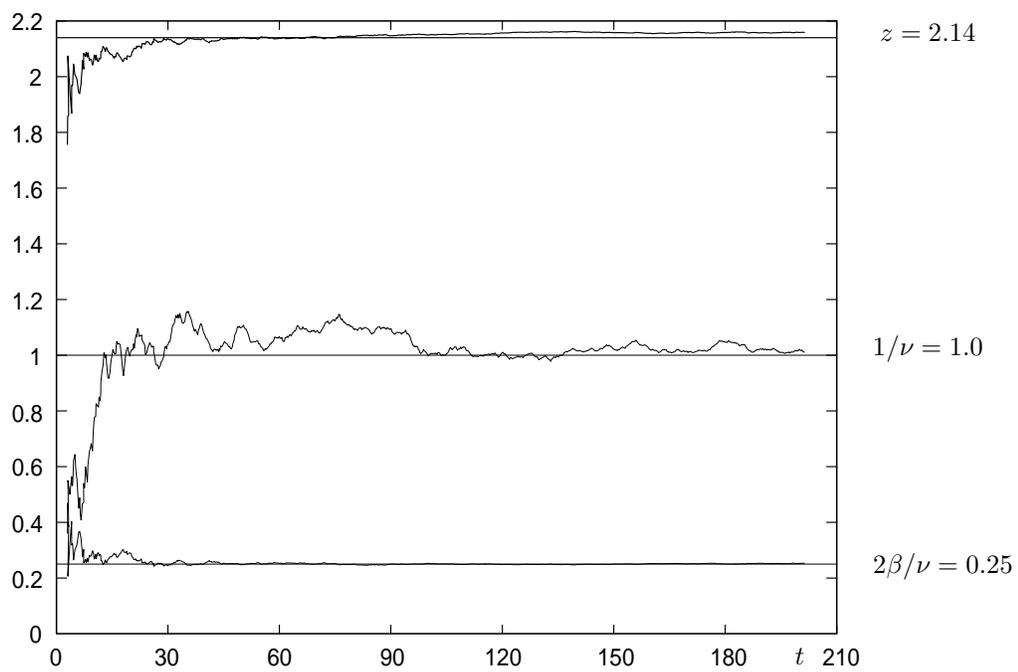}}}
\end{picture}
\caption{\footnotesize
	The curves show the values for $z$, $2\beta/\nu$ and $1/\nu$
	calculated for initial magnetization $m_0=1$
        with the local scaling fit.
}
\label{dd1_stp}
\end{figure}

In order to have more rigorous understanding of the dynamic
scaling,
we alternatively present a {\it local} approach for estimating
the critical exponents, in contrast to the {\it global}
scaling fitting procedure discussed above.
For example, comparing the functions
$U(t,L)$ and
$U(t',2L)$,  for each time step $t$ we search for
$t'$ such that $U(t,L)=U(t',2L)$,
and from the ratio
$t'/t=2^z$ the value of $z$ is obtained according to
Eq.(\ref{e4}). For the same time step $t$
this particular value of  $z$
can be used to estimate $2\beta/\nu$
from Eq.(\ref{e5}) and  $1/\nu$
from Eq.(\ref{es}). Then we obtained all the exponents
as functions of the time $t$.
The result is shown in Fig.~\ref{dd1_stp} for a pair
of lattices with $L=32$ and $L=64$. In order to guide the eye,
three horizontal lines $z=2.14$, $2\beta/\nu=0.25$, and $1/\nu=1$
are included,
the latter $2$ values being the exact results for the 2-dimensional
Ising model.
The figure shows clearly that the fluctuations
for smaller time $t$ are big. But for $z$ and in particular
for $2\beta/\nu$ the curve tends very nicely to a
horizontal line for $t\gsim30$.
The situation is less satisfactory for the curve of
$1/\nu$ where it shows still some fluctuations up to
a fairly late time $t$. The reason
may be either less statistics or from the approximation of
the differentiation by a difference.
The exponents $z$, $2\beta/\nu$ and $1/\nu$ can be obtained by
averaging over the time direction.
To show the effect of the
fluctuations for smaller $t$, one takes the average  starting
from different initial times.
The values of the exponents obtained
in this way have been given in Tab.~\ref{t2}.
The errors have again been estimated
by a comparison of each run for $L$ and with each
run for  $L'=2L$.

In Fig.~\ref{dd1_stp}, when $t\lsim30$ the exponent $z$ is somewhat
small.
This might be due to the effect of $t_{mic}$.
In some cases, e.g. in the measurement of $\theta$
\cite{li94,sch95,oka95} and $\beta/\nu$ in the next section,
this effect hardly shows up. However, in some other cases
it can remain till $t\sim20-30$ \cite{sch95,oka95}.
This kind of effect probably comes from the fact that the initial
magnetization density is not uniform enough.
As we have already mentioned before, from Fig.~\ref{dd1_stp}
one can see explicitly even for $t\gsim30$  that
the exponent $z$ slightly rises as time
evolves. Even though the reason is not clear, but it is
interesting that the tendency  of $z$ to rise
does not affect the measurement
of the static exponents, especially
$2\beta/\nu$. They are quite stable.
On the other hand, even for $z$ itself,
$1\%$ of the fluctuation in the time direction should
also be not too bad.
It may be due to the finite size effect or some technical reasons.

The curves from a scaling fit of the
lattices $8\leftrightarrow16$ and $16\leftrightarrow32$
which are not shown in the figures here
look qualitatively the same as those in Fig.~\ref{dd1_stp}, but
the fluctuations for $1/\nu$ are somewhat larger.
The curve
for $z$ from the smaller lattice sizes $8\leftrightarrow16$, however,
rises
continuously for $t\gsim20$, thus showing that such a lattice size
is too small for this kind of analysis.
The result of an average of these values are also included in
Tab.~\ref{t2}.

\vspace{0.3cm}

To measure $z$, $2\beta/\nu$ and $1/\nu$,
instead of $U$, $M^{(2)}$ and $\partial_\tau M^{(2)}$  one
can also use $\tilde U$, $|M|$ and $\partial_\tau |M|$ with
\be
\tilde U(t,L) =1-\frac{  M^{(2)} }{ |M|^2 }.
\label{ezs}
\ee
{}From Eq.(\ref{e2}) one easily deduces that the relation (\ref{e4})
holds
for $U$ as well as for $\tilde U$, and (\ref{es}) holds as well for
$\partial_\tau \ln|M|$.
Only Eq.(\ref{e5}) is slightly modified to
\be
|M(t,L)| = b^{\beta/\nu} |M(b^{z}t,L')|; \qquad L'=bL,
\label{e5s}
\ee
We do not plot the curves here,
since they look very much the same as Figs.\ref{ddU}-\ref{ddd} for
the
global and local analysis.
 For simplification
we have  compared only the two lattice sizes $L=32$ and $L=64$,
where we have used the full time scale up to $t'_{max}=900$ for
$L=64$.
The results for the global fit
have been included in the lower part of
Tab.~\ref{t1}.
All the values in the table are remarkably
consistent.
The same holds for
the results from the local approach. They are shown in Tab.~\ref{t2}.

\section{The Critical Relaxation \protect\newline with
         Initial Magnetization One}

In the previous section we have investigated
the finite size scaling of the critical relaxation of the Ising model
up to even the macroscopic short-time scale, starting from
a random state with zero initial magnetization,
i.e., a completely disordered state.
{}From a measurement of
the time evolution of the observables $|M(t,L)|$, $M^{(2)}(t,L)$,
and $M^{(4)}(t,L)$ together with the scaling relation (\ref{e2})
 the critical exponents $z$, $2\beta/\nu$
and $1/\nu$ were obtained. They are in good agreement with the known
results. This is a strong support for the scaling relations
derived by Janssen, Schaub and Schmittmann \cite{jan89}.

At this stage one may ask whether there is also
a scaling relation for the critical relaxation starting from
a completely ordered state, i.e. with initial magnetization
$m_0=1$. It has been known for some time that, before the
exponential decay of the magnetization starts,
there exists a time regime where the magnetization
behaves non-linearly and decays according to a power law.
The question is only {\it when} such a scaling behaviour starts.
Some effort has been made in this direction \cite{sta92,mun93}
with Monte Carlo simulation.
The authors have
simulated the critical relaxation with an extreme big lattice
but only up to a quite short evolution time and
have estimated the exponent
$\beta/{\nu z}$ from the power law decay of M(t). However, the result
has not been so clear and also other exponents as $z$ and $1/\nu$
have not been obtained.

%############################## BEGINN  TAB3 #######################
\begin{table}[h]\centering
$$
\begin{array}{|c|c|c|c|l|l|l|}
\hline
\mbox{Input} & \mbox{Lattice} &  t'_{min} & t'_{max} &\quad z\ &
\quad 2\beta/\nu\ &\quad
1/\nu\  \\
\hline
\hline
          &               & \ 50 & 300 & 2.121(4) & 0.2489(4) &
1.03(2)  \\
\cline{5-7}
  U       &   32\leftrightarrow 64 &  200 &  &2.122(5) & 0.2491(5) &
1.02(2)  \\
\cline{3-7}
 M^2         &                  & \ 50 & 900 & 2.129(5) & 0.2503(5) &
1.04(2)  \\
 \cline{5-7}
       &                      &  200 & & 2.129(5) & 0.2505(6) &
1.04(2)  \\\cline{3-6}
\hline
\hline
 \tilde U  & 32\leftrightarrow 64 & \ 50 & 900& 2.140(5) & 0.2514(6)&
1.07(2)  \\
\cline{5-7}
|M|        &                      &  200 & & 2.141(5) & 0.2515(7) &
1.07(2)  \\
\hline
\end{array}
$$
\caption{
\footnotesize
 Results for $z$, $2\beta/\nu$ and $1/\nu$, respectively,
from the 2-dimensional Ising model with initial magnetization
$m_0=1$.
The values are
obtained from a global scaling fit for two lattices.
}
\label{t3}
\end{table}
%############################## ENDE  TAB3 #######################

In this section we study systematically the
scaling behaviour of the critical relaxation from a completely
ordered initial state, but in {\it finite systems },
following a procedure parallel to that discussed in the previous
section.
The advantage is that a not too big finite system allows
for more statistics and longer evolution time even though the
power law decay of the magnetization will not be perfect.
{}From our results we confirm that the scaling appears in a quite
early stage of the relaxation as in the case of $m_0=0$.

Here we will only present data
from the scaling collapse for lattice sizes $L=32$ and $L=64$.
Also we confine ourselves to 4 runs with 50\,000 updations each,
instead of 8 runs in the previous section.

\begin{figure}[p]\centering
\epsfysize=12cm
\epsfclipoff
\fboxsep=0pt
\setlength{\unitlength}{1cm}
\begin{picture}(13.6,12)(0,0)
\put(0.2,8.2){\makebox(0,0){$U(t,L)$}}
\put(8.2,4.1){\makebox(0,0){\footnotesize $L=64$}}
\put(5.2,1.7){\makebox(0,0){\footnotesize $L=32$}}
\put(10.7,.3){\makebox(0,0){$t$}}
\put(0,0){{\epsffile{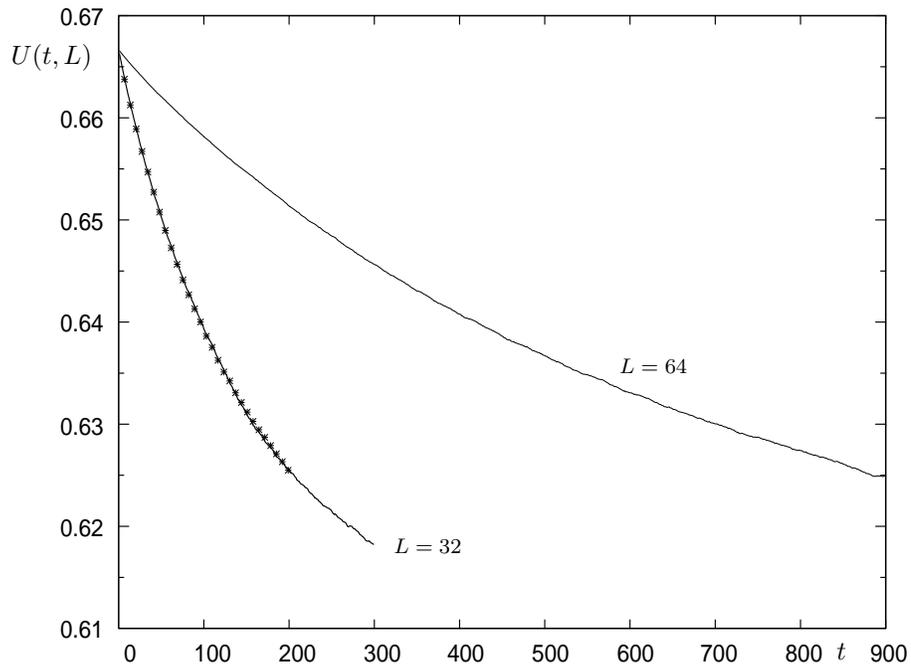}}}
\end{picture}
\caption{\footnotesize
   The cumulants $U(t,L)$ for $L=32$ and $64$
   for initial magnetization $m_0=1$
   plotted
	versus time $t$. The dots fitted to the line of
         $L=32$ show
	the cumulant with lattice size $2L$ rescaled in time
        by the best fit value $2^z$ given in Table~3.
}
\label{ddUm1}
\end{figure}

\begin{figure}[p]\centering
\epsfysize=12cm
\epsfclipoff
\fboxsep=0pt
\setlength{\unitlength}{1cm}
\begin{picture}(13.6,12)(0,0)
\put(0.0,8.2){\makebox(0,0){$M^{(2)}(t,L)$}}
\put(9.0,2.7){\makebox(0,0){\footnotesize $L=32$}}
\put(9.4,1.8){\makebox(0,0){\footnotesize $L=64$}}
\put(10.8,.3){\makebox(0,0){$t$}}
\put(0,0){{\epsffile{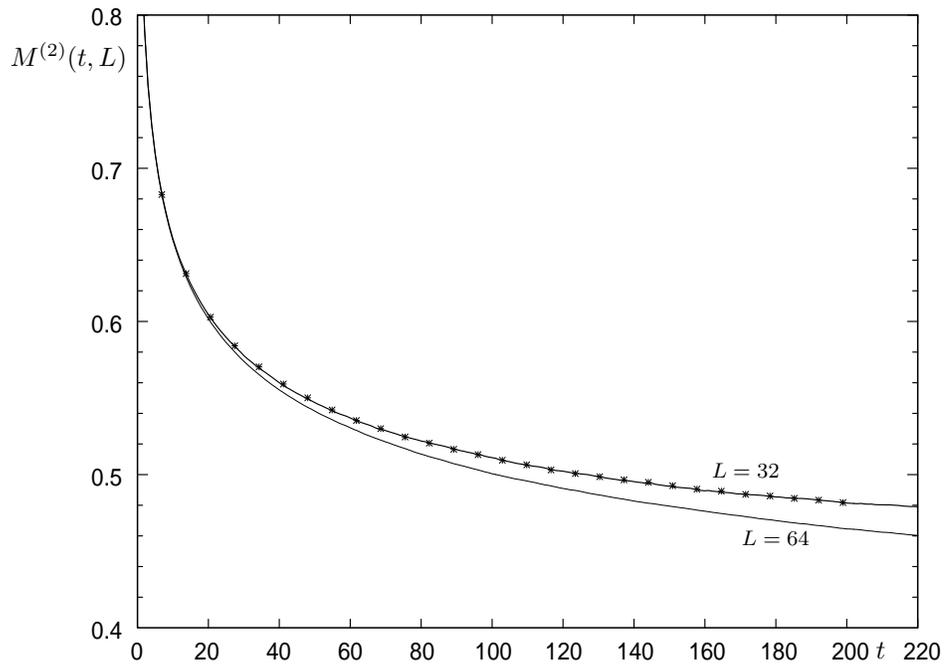}}}
\end{picture}
\caption{\footnotesize
   The second moments $M^{(2)}$ for $L=32$ and $64$
   for initial magnetization $m_0=1$
   plotted
	versus time $t$. The dots fitted to the curve for $L=32$
	show the second moment with lattice size $L=64$ rescaled in
	 time by the best fit value $2^z$ and
	$2\beta/\nu$ given in Table~3.
}
\label{ddMm1}
\end{figure}

\begin{figure}[p]\centering
\epsfysize=12cm
\epsfclipoff
\fboxsep=0pt
\setlength{\unitlength}{1cm}
\begin{picture}(13.6,12)(0,0)
\put(-.4,8.1){\makebox(0,0){$\partial_\tau\ln M^{(2)}(t,L)$}}
\put(9.0,6.4){\makebox(0,0){\footnotesize $L=32$}}
\put(9.4,8.2){\makebox(0,0){\footnotesize $L=64$}}
\put(10.8,.3){\makebox(0,0){$t$}}
\put(0,0){{\epsffile{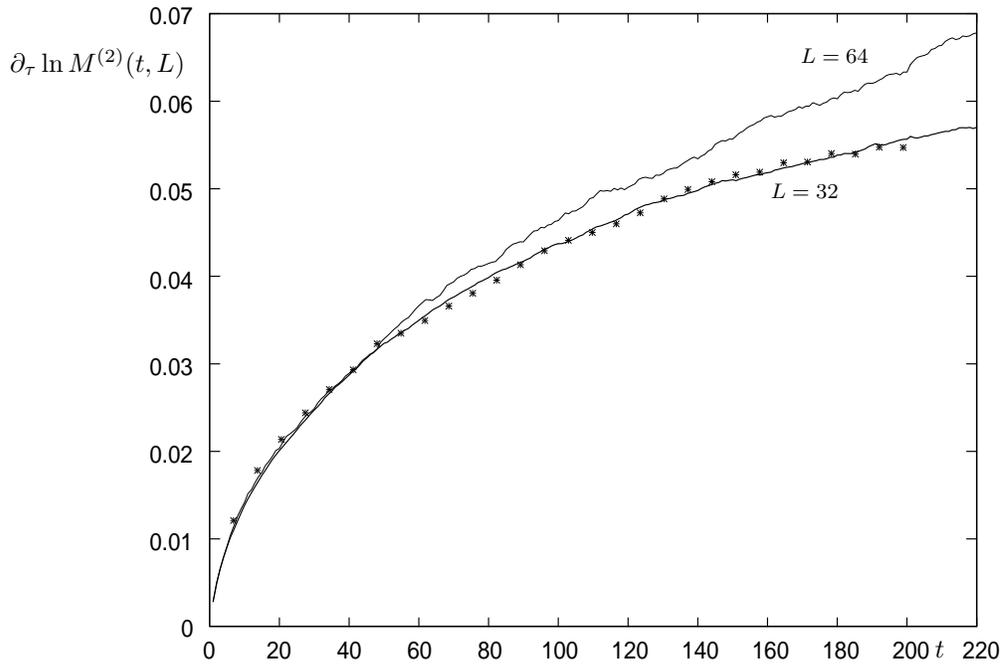}}}
\end{picture}
\caption{\footnotesize
       $\partial_\tau \ln M^{(2)}(t,L)$ for $L=32$
	and $64$
   for initial magnetization $m_0=1$
	plotted
	versus time $t$. The dots fitted to  the line for $L=32$
	show that with lattice size $L=64$ rescaled in time by
        the best fit values $2^z$
	and $b^{1/\nu}$ given in Table~3.
}
\label{dddm1}
\end{figure}

\begin{figure}[p]\centering
\epsfysize=12cm
\epsfclipoff
\fboxsep=0pt
\setlength{\unitlength}{1cm}
\begin{picture}(13.6,12)(0,0)
\put(12.5,8.6){\makebox(0,0){$z=2.14$}}
\put(12.5,4.3){\makebox(0,0){$1/\nu=1.0$}}
\put(12.7,1.5){\makebox(0,0){$2\beta/\nu=0.25$}}
\put(10.8,.3){\makebox(0,0){$t$}}
\put(0,0){{\epsffile{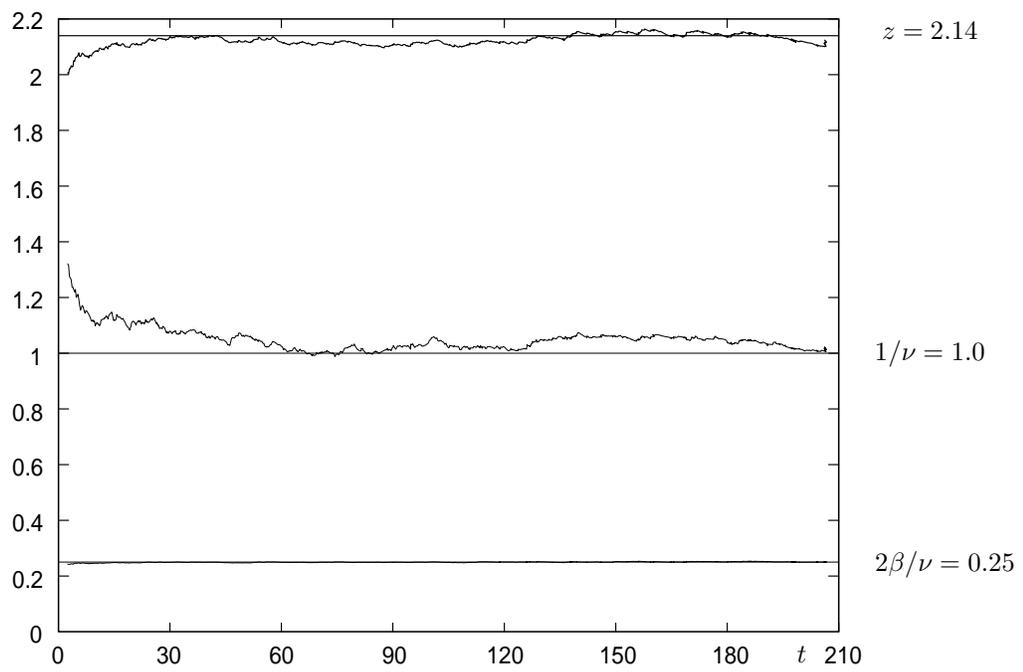}}}
\end{picture}
\caption{\footnotesize
The curves show the values of $z$, $2\beta/\nu$ and $1/\nu$ obtained
for initial magnetization $m_0=1$
from a local scaling fit for $L=32$ and $L=64$.
}
\label{dd_m1_stp}
\end{figure}

The curves for the second moment of
the magnetization start with the value one
for $t=0$ and
decrease for later time.
This is seen in Fig.~\ref{ddMm1}.
A similar decrease is found for the
curves for the cumulants $U(t,L)$ in Fig.~\ref{ddUm1},
while $\partial_\tau \ln M^{(2)}(t,L)$ in
Fig~\ref{dddm1} shows a rising behaviour. In all three
figures points mark the values
for $L=64$ rescaled in time by the best fit values
of $z$, $2\beta/\nu$, and
$1/\nu$. Surprisingly here we also observe very nice dynamic scaling.
 Table~\ref{t3}
shows the results of the global fitting procedure
up to $t'_{max}=300$ or
$t'_{max}=900$, respectively.
Since the values of $U$, $M^{(2)}$ and $\partial_\tau M^{(2)}$
  for smaller $t$ are large now compared with those in the case of
$m_0=0$ in the previous  section, the results show
less fluctuations for smaller time $t$.
Actually one may also expect that due to the unique
initial configuration less statistics is needed
to obtain stable results.
This is also supported
by Fig.~\ref{dd_m1_stp} from the local approach.
It is very interesting that the exponent $2\beta/\nu$
shows almost {\it invisible } fluctuations
in the whole time regime even up to the very beginning,
although
$z$ has some similar unstable behaviour as that in the case of
$m_0=0$. Especially in the first $30$ time steps its value is also
a bit small. The fact that in the first steps of $t$
the exponent $z$ is quite near to $2.0$ might indicate that
at the very beginning of the time evolution the system
is ``classical''.
 Similarly as Tab.~\ref{t2} of the previous section,
Tab.~\ref{t4} gives the
averages over the time direction, starting
at different initial values $t'_{min}$.

%############################## BEGINN  TAB4 #####################
\begin{table}[h]\centering
$$
\begin{array}{|c|c|c|l|l|l|}
\hline
\mbox{Input} & \mbox{Lattice} &  t'_{min}   &\quad z\ & \quad
2\beta/\nu\ &\quad
1/\nu\  \\
\hline
\hline
 U        & 32\leftrightarrow 64 & \ 50 &  2.122(7) & 0.2508(3) &
1.04(2) \\
\cline{3-6}
M^2
       &                      &  200 &  2.122(8) & 0.2510(4) &
1.04(2) \\
\hline
\hline

 \tilde U  & 32\leftrightarrow 64 & \ 50 &  2.133(6) & 0.2516(4) &
1.06(2)  \\
\cline{3-6}

|M|        &                      &  200 &  2.134(7) & 0.2520(5) &
1.06(3)  \\
\hline
\end{array}
$$
\caption{
\footnotesize
 Results for $z$, $2\beta/\nu$ and $1/\nu$, respectively,
from the $2$-dimensional Ising model with initial magnetization
$m_0=1$.
Values are obtained from the average in the time direction
from $t'_{min}$ up to $t'_{max}=900$
with a local scaling fit.
}
\label{t4}
\end{table}
%############################## ENDE  TAB4 ####################

As in the previous section, we carry out also the analysis
with $\tilde U$ defined in (\ref{ezs}),
$|M|$ and $\partial_\tau|M|$.
 The results have been included in the lower parts
of Tab.~\ref{t3} from the global fit,
 or in Tab.~\ref{t4} from the local approach, respectively.
They again have a similar quality as those
reported above. In comparison with the results in
\cite{sta92,mun93,dam93}, the value of $z$ we obtained
is definitely smaller and near to the one from
the $\epsilon$-expansion and other traditional
measurements \cite{wil85,tan87}.
In case of $m_0=0$, the results for $z$ measured
from $U$ and $\tilde U$ are almost the same.
However, in case of $m_0=1$, the value measured
from $U$ is somewhat smaller than that measured from $\tilde U$.
In the construction of $U$ we have not subtracted the
odd moments. This may have some effect on the measurement of $z$.

As compared to those in the previous  section for
the case of $m_0=0$, the results here are somewhat more stable.
This may really indicate that it is also promising
to measure the critical exponents from the critical
relaxation process starting from
an ordered state
even though some more theoretical arguments like that by
Janssen, Schaub and Schmittmann \cite{jan89} are still needed.
It is clear that in case of a random initial state
with $m_0=0$ all the observables
discussed start their evolution from zero and therefore
the fluctuations at the beginning of the time evolution
 are natural bigger.
Besides this, the effect of the non-uniformity of the
magnetization density
in the practically
generated initial configurations is not completely
negligible for not too big lattices.
Another way to measure the critical exponents
 is to study the critical relaxation
starting from an initial state with small but non-zero
initial magnetization \cite {sch95,oka95}.
However, the situation is not so clear
 since the new dynamic exponent $\theta$
enters the calculation. Further investigation is needed.

\begin{figure}[p]\centering
\epsfysize=12cm
\epsfclipoff
\fboxsep=0pt
\setlength{\unitlength}{1cm}
\begin{picture}(13.6,12)(0,0)
\put( .3,7.1){\makebox(0,0){\footnotesize $\ln M(t)$}}
\put(7.4,5.4){\makebox(0,0){\footnotesize (b) $\ln M(t)$ vs.
$c\ln(t)$}}
\put(4.4,3.2){\makebox(0,0){\footnotesize (a) $\ln M(t)$ vs. $t$}}
\put(0,0){{\epsffile{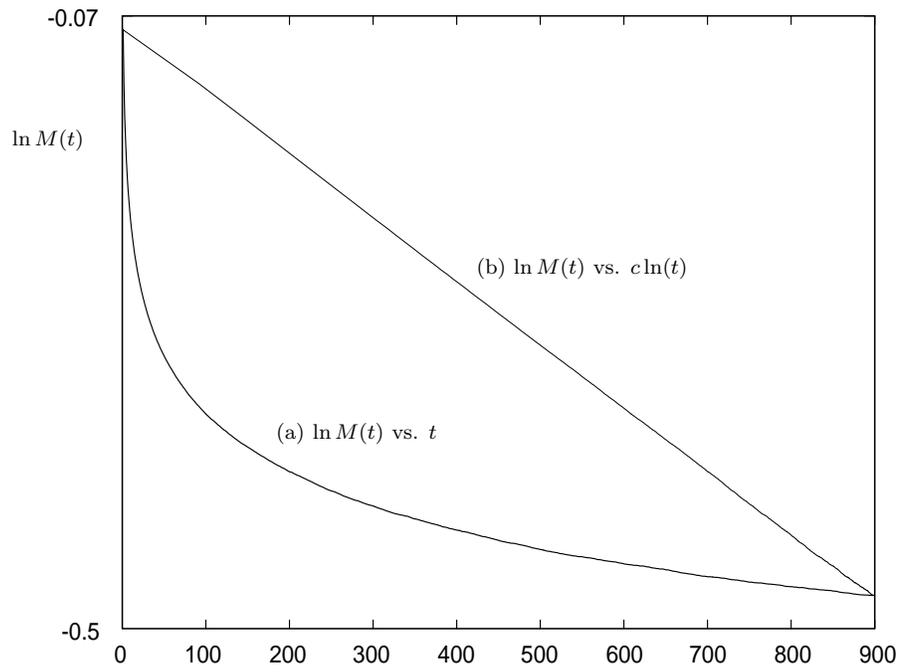}}}
\end{picture}
\caption{\footnotesize
The quantities $\ln M(t,L)$ for $L=64$ and $m_0=1$
plotted (a) versus $t$, and (b) versus $c\ln(t)$.
The factor $c$ has been chosen such that at $900$
of the abscissa both curves coincide.
}
\label{t3a}
\end{figure}

\begin{figure}[p]\centering
\epsfysize=12cm
\epsfclipoff
\fboxsep=0pt
\setlength{\unitlength}{1cm}
\begin{picture}(13.6,12)(0,0)
\put(12.3,6.7){\makebox(0,0){\footnotesize $z=2.14$}}
\put(0.4,8.3){\makebox(0,0){\footnotesize $z$}}
\put(10.8,.2){\makebox(0,0){\footnotesize $t$}}
\put(0,0){{\epsffile{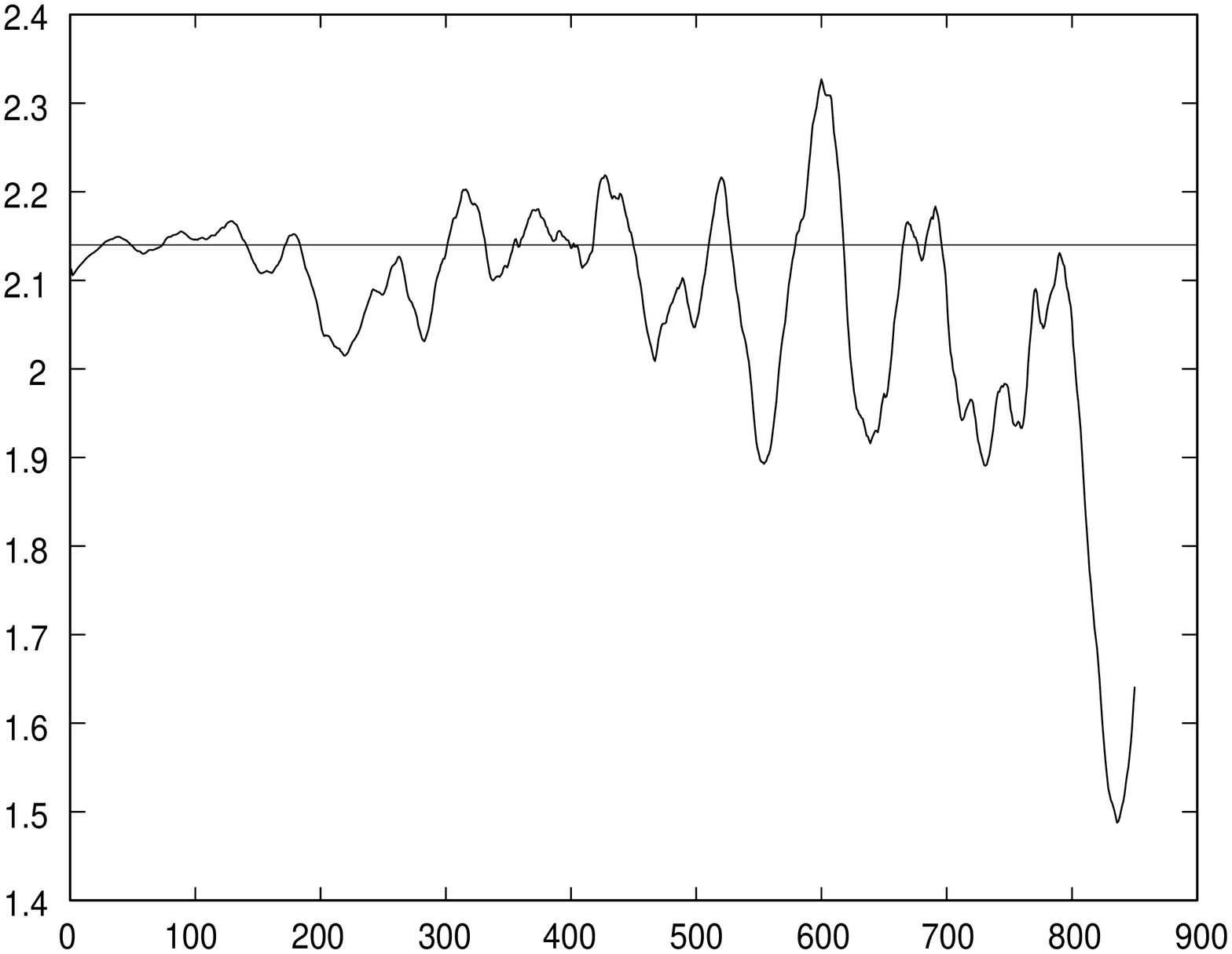}}}
\end{picture}
\caption{\footnotesize
The exponent $z$ calculated from the slope of the magnetization
$M(t)$ in a double-log scale.
The slope $\beta/(\nu z)$ has been fitted for each time step
within the time interval $[t,t+50]$.
In order to calculate $z$ we have used $\beta/\nu=1/8$ as input.
}
\label{f10}
\end{figure}

Finally we plot the time evolution of the magnetization
with both double-log scale and semi-log scale in order to see
whether it has entered the regime of linear decay  or not.
In Fig.~\ref{t3a} the straight line shows definitely
that the magnetization is still in the regime of
non-linear decay.
{}From the slope of $M(t)$ in double-log scale one can obtain the
exponent
$\beta/(\nu z)$. for each time $t$ we have measured it by a least
square fit
within a time interval $[t,t+50]$. In Fig.~\ref{f10}
instead of $\beta/(\nu z)$ the exponent $z$ is plotted vs. time using
the exact value $\beta/\nu=1/8$. Note that the time scale in
Fig.~\ref{f10} is different from that in
Fig.~\ref{dd_m1_stp}.
Figure~\ref{f10} shows that in the regime $150 \gsim t \gsim
t_{mic}\approx30$
the values for $z$ are rather consistent with those obtained before,
especially those measured from $\tilde U$ in Tab.~\ref{t3}.
However, the lattice size $L=64$ seems not to be big enough to
present
a rigorous power law behaviour in the whole time region.

\section{Discussion}

 We numerically simulate the critical relaxation process of the
two-dimensional Ising model with the initial state
both completely disordered or completely ordered.
 Based on the finite size scaling
for the dynamics at the early time,
both the static and dynamic critical exponents
are measured. To determine $z$ independently,
a time-dependent Binder cumulant is constructed.
The value of $z$ measured from the critical relaxation
from a completely ordered state is slightly smaller
than that from a completely disordered state.
The reason is not yet clear.
Taking
the average of the four measurements of $z$ from the global scaling
fit
of $U$ and $\tilde U$ within the time interval $[50,900]$ of $t'$
for both relaxation processes, see Tab.~\ref{t1} and Tab.~\ref{t3},
we conclude

\vspace*{0.3cm}
\centerline{$z = 2.143(5)$}
\vspace{0.2cm}

\noindent This should be compared with the existing numerical
results $z=2.13(8)$
from \cite{wil85}, and $z=2.14(5)$ from \cite{tan87},
 and also with $z=2.126$ obtained from an
$\epsilon$-expansion in \cite{bau81}, even though some bigger
values
are also reported recently \cite{sta92,mun93,dam93}.
It is remarkable that from the short-time dynamics
one can not only efficiently measure the dynamic
exponent $z$, but also the static exponents.
Especially the quality of the exponent
$2\beta/\nu$ is very good.
All these results provide strong confirmation for the scaling
relation at the early time of the critical relaxation process.
Compared with the traditional methods,
the advantage of our {\it dynamic } Monte Carlo algorithm is
that the measurement is carried out in the beginning of the time
evolution rather than in the equilibrium where
critical slowing down is more severe. Therefore our method is
efficient. Compared with the
non-local algorithms, our dynamic algorithm can study
the properties of the original local dynamics.
On the other hand, it has recently been suggested that
the critical exponents
can also be measured from the {\it power law} behaviour
of the observables including the auto-correlation
in the macroscopic short-time regime in a large enough
lattice \cite {sch95,oka95}. Compared with that approach,
the advantage of estimating the exponents
from the dynamic finite size scaling
as reported in this paper is that one needs not too big lattices.
However, the result has
to be obtained by comparing two
lattices and longer time of the evolution for the bigger lattice
should
be carried out.

It is somewhat surprising that for the critical relaxation
from the completely ordered initial state
there exist also universality and scaling in such
an early stage of the time evolution. Further investigation
especially on a more general critical relaxation process
from an ordered state with initial magnetization $m_0$
smaller but near to one can be interesting. One may expect that a
new
dynamic exponent should be introduced in order to complete
the scaling relation.

\vspace{0.5cm}

{\bf Acknowledgement:} One of the authors (Z.B.L.)
    is grateful to the Alexander von
Humboldt-Stiftung for a fellowship.
The authors would like to thank
   K. Untch for the help in maintaining our Workstations.

\end{document}